    \documentstyle[12pt]{article}
\begin{document}
\baselineskip=14pt plus 1pt minus 1pt

\begin{center}
{\Large \bf  Rotations of Nuclei with Reflection Asymmetry
Correlations}
\end{center}
\medskip

\begin{center}
{\large Nikolay Minkov$^*$\footnote[1] {e-mail:
nminkov@inrne.bas.bg}, S. B. Drenska$^*$, P. P. Raychev$^{*}$, R.
P. Roussev$^*$, and Dennis Bonatsos$^\dagger$  \\
\medskip
$^*$  Institute for Nuclear Research and Nuclear Energy, \\
72 Tzarigrad Road, 1784 Sofia, Bulgaria\\
\medskip
$^\dagger$ Institute of Nuclear Physics, N.C.S.R. ``Demokritos'',\\
GR-15310 Aghia Paraskevi, Attiki, Greece}
\end{center}
\bigskip

\begin{abstract}
We propose a collective Hamiltonian which incorporates interactions
capable to generate rotations in nuclei with simultaneous presence
of octupole and quadrupole deformations. It is demonstrated that
the model formalism could be applied to reproduce the staggering
effects observed in nuclear octupole bands. On this basis we
propose that the interactions involved would provide a relevant
handle in the study of collective phenomena in nuclei and other
quantum mechanical systems with reflection asymmetry correlations.
\end{abstract}
\bigskip

The properties of nuclear systems with octupole deformations
\cite{BM75} are of current interest due to the increasing number of
evidences for the presence of octupole instability in different
regions of nuclear table \cite{BN96}. Various parametrizations of the
octupole degrees of freedom have opened a useful tool for
understanding the role of the reflection asymmetry correlations and
for analysis of the collective properties of such kind of systems
\cite{Ro90,Ham91,WD99}. As an important step in this direction it is
necessary to elucidate the question: which are the collective nuclear
interactions that correspond to the different octupole shapes and how
do they determine the structure of the respective energy spectra?
The physically meaningful answer could be obtained by taking
into account the simultaneous presence of other collective degrees of
freedom, such as the quadrupole ones.

In the present work we address the above problem by examining the
interactions that generate collective rotations in a system with
octupole deformations. Based on the octahedron point symmetry
parametrization of the octupole shape \cite{Ham91}, we propose a
general collective Hamiltonian which incorporates the interactions
responsible for the rotations associated with the different octupole
deformations. It will be shown that after taking into account the
quadrupole degrees of freedom and the appropriate higher order
quadrupole-octupole interaction the model formalism would be able to
reproduce schematically some interesting effects of the fine
rotational structure of nuclear octupole bands. The study is strongly
motivated by the need of theoretical explanation of the recently
observed staggering patterns in octupole bands of light actinides
\cite{DBoct00} as well as by the possibility to gain an insight into
the fine structure of negative parity rotational bands based on
octupole vibrations.

Our model formalism is based on the understanding that the
collective properties of a physical system in which octupole
correlations take place should be influenced by the following
most general octupole field $V_{3}=\sum_{\mu =-3}^{3}
\alpha_{3\,\mu}^{fix}Y^{*}_{3\,\mu}$, (in the intrinsic,
body-fixed frame) which can be written in the form \cite{Ham91}:
\begin{equation}
V_{3}=\epsilon_{0}A_{2}+\sum_{i=1}^{3}\epsilon_{1}(i)F_{1}(i)+
\sum_{i=1}^{3}\epsilon_{2}(i)F_{2}(i) \ ,
\label{field2}
\end{equation}
where the quantities
\begin{eqnarray}
A_{2}&=&-\frac{i}{\sqrt{2}}(Y_{3\, 2}-Y_{3\, -2})
     =\frac{1}{r^{3}}\sqrt{\frac{105}{4\pi}}xyz \ ,
     \label{A2} \\
F_{1}(1)&=&Y_{3\, 0}=\frac{1}{r^{3}}\sqrt{\frac{7}{4\pi}}
           z(z^{2}-\frac{3}{2}x^{2}-\frac{3}{2}y^{2}) \ , \\
F_{1}(2)&=&-\frac{1}{4}\sqrt{5}(Y_{3\, 3}-Y_{3\, -3})
+\frac{1}{4}\sqrt{3}(Y_{3\, 1}-Y_{3\, -1}) \nonumber \\
        &=&\frac{1}{r^{3}}\sqrt{\frac{7}{4\pi}}
           x(x^{2}-\frac{3}{2}y^{2}-\frac{3}{2}z^{2}) \ , \\
F_{1}(3)&=&-i\frac{1}{4}\sqrt{5}(Y_{3\, 3}+Y_{3\, -3})
-i\frac{1}{4}\sqrt{3}(Y_{3\, 1}+Y_{3\, -1}) \nonumber \\
        &=&\frac{1}{r^{3}}\sqrt{\frac{7}{4\pi}}
           y(y^{2}-\frac{3}{2}z^{2}-\frac{3}{2}x^{2})\ , \\
F_{2}(1)&=&\frac{1}{\sqrt{2}}(Y_{3\, 2}+Y_{3\, -2})
         =\frac{1}{r^{3}}\sqrt{\frac{105}{16\pi}}
          z(x^{2}-y^{2})\ , \\
F_{2}(2)&=&\frac{1}{4}\sqrt{3}(Y_{3\, 3}-Y_{3\, -3})
+\frac{1}{4}\sqrt{5}(Y_{3\, 1}-Y_{3\, -1}) \nonumber  \\
        &=&\frac{1}{r^{3}}\sqrt{\frac{105}{16\pi}}
          x(y^{2}-z^{2})\ , \\
F_{2}(3)&=&-i\frac{1}{4}\sqrt{3}(Y_{3\, 3}+Y_{3\, -3})
+i\frac{1}{4}\sqrt{5}(Y_{3\, 1}+Y_{3\, -1}) \nonumber  \\
        &=&\frac{1}{r^{3}}\sqrt{\frac{105}{16\pi}}
          y(z^{2}-x^{2}), \label{F23}
\end{eqnarray}
(with $r^{2}=x^{2}+y^{2}+z^{2}$) belong to the irreducible
representations (irreps) of the octahedron group ($O$). $A_{2}$
is one-dimensional, while $F_{1}$ and $F_{2}$ are
three-dimensional irreps. The seven real parameters $\epsilon_{0}$
and $\epsilon_{r}(i)$ ($r=1,2;\, i=1,2,3$) determine the
amplitudes of the octupole deformation. Their relation to the
$\alpha_{3\,\mu}^{fix}$ is given in \cite{Ham91}.

Our proposition is that the general collective Hamiltonian which
incorporates the shape characteristics of the octupole field
(\ref{field2}) can be constructed on the basis of the above
octahedron irreps. For this purpose we introduce operator forms
of the quantities $A_{2}$, $F_{1}(i)$ and $F_{2}(i)$ ($i=1,2,3$)
in which the cubic terms of the Cartesian variables $x$, $y$ and
$z$ in Eqs~(\ref{A2})--(\ref{F23}) are replaced by appropriately
symmetrized combinations of cubic terms of the respective angular
momentum operators $\hat{I}_{x}$, $\hat{I}_{y}$, $\hat{I}_{z}$
(with $\hat{I}^{2}=\hat{I}_x^{2}+\hat{I}_y^{2}+\hat{I}_z^{2}$).
The following Hamiltonian is then obtained:
\begin{equation}
\hat{H}_{oct}=\hat{H}_{A_{2}}+
\sum_{r=1}^{2}\sum_{i=1}^{3}\hat{H}_{F_{r}(i)} \ ,
\label{Hoctgen}
\end{equation}
with
\begin{eqnarray}
\hat{H}_{A_{2}}&=&{a}_{2}\frac{1}{4}
[(\hat{I}_x\hat{I}_y+\hat{I}_y\hat{I}_x)\hat{I}_z+
\hat{I}_z(\hat{I}_x\hat{I}_y+\hat{I}_y\hat{I}_x)] \ ,
\label{HA} \\
\hat{H}_{F_{1}(1)}&=&\frac{1}{2}{f}_{11}
\hat{I}_z(5\hat{I}_z^{2}-3\hat{I}^{2}) \ ,
\label{HF11} \\
\hat{H}_{F_{1}(2)}&=&\frac{1}{2}{f}_{12}
(5\hat{I}_x^{3}-3\hat{I}_x\hat{I}^{2}) \ ,
\label{HF12} \\
\hat{H}_{F_{1}(3)}&=&\frac{1}{2}{f}_{13}
(5\hat{I}_y^{3}-3\hat{I}_y\hat{I}^{2}) \ ,
\label{HF13} \\
\hat{H}_{F_{2}(1)}&=&{f}_{21}\frac{1}{2}
[\hat{I}_z(\hat{I}_x^{2}-\hat{I}_y^{2})+
(\hat{I}_x^{2}-\hat{I}_y^{2})\hat{I}_z] \ ,
\label{HF21} \\
\hat{H}_{F_2(2)}&=&{f}_{22}
(\hat{I}_x\hat{I}^{2}-\hat{I}_x^{3}-
\hat{I}_x\hat{I}_z^{2}-\hat{I}_z^{2}\hat{I}_x) \ ,
\label{HF22} \\
\hat{H}_{F_2(3)}&=&{f}_{23}
(\hat{I}_y\hat{I}_z^{2}+\hat{I}_z^{2}\hat{I}_y+
\hat{I}_y^{3}-\hat{I}_y\hat{I}^{2})
\label{HF23}
\end{eqnarray}
The Hamiltonian parameters ${a}_{2}$ and ${f}_{r\, i}$ ($r=1,2;\,
i=1,2,3$) are formally related to the parameters in
(\ref{field2}) as follows ${a}_{2}=\epsilon_{0}\sqrt{105/(4\pi)}$,
${f}_{1\, i}=\epsilon_{1}(i)\sqrt{7/(4\pi)}$, ${f}_{2\,
i}=\epsilon_{2}(i)\sqrt{105/(16\pi)}$, $i=1,2,3$.

During the procedure described above, the $r^3$ factors appearing
in the denominators of Eqs~(\ref{A2})--(\ref{F23}) are replaced by
$\hat I^3$ factors. In the final result,
Eqs~(\ref{HA})--(\ref{HF23}), we normalize with respect to $\hat
I^3$, i.e. we multiply the results by $\hat I^3$, an operation
which is equivalent to the transition to a unit sphere, a natural
thing to do since we are interested in surface shapes.

We remark that the terms of the Hamiltonian obtained (as a
function of the angular momentum operators $\hat I_x$, $\hat
I_y$, $\hat I_z$) correspond to the same octupole shapes which
appear in Eqs~(\ref{A2})--(\ref{F23}) and belong to the same
irreps of the octahedron group.  In other words, through the
above procedure we determine the octahedron point symmetry
properties of the system in angular momentum space.

Our analysis shows that the operator $\hat{H}_{F_{1}(1)}$,
Eq.~(\ref{HF11}), which corresponds to $Y_{3\, 0}$ (with axial
deformation) is the only one octupole operator possessing diagonal
matrix elements in the states with collective angular momentum
$I$. Below it will be shown that it is of major importance for
determining the fine structure of collective bands with octupole
correlations. Actually, it is well known that the $Y_{3\, 0}$
(axial) deformation is the leading mode in the systems with
reflection asymmetric shape (See for review \cite{BN96}).

Further, it is known that the use of the pure octupole field
(\ref{field2}) is not sufficient to incorporate the collective
shape properties of the system. More specifically a unique
parametrization of the pure  octupole field  in an intrinsic
frame has not been obtained yet in a consistent way \cite{BN96}.
In this respect the consideration of octupole degrees of freedom
together with the quadrupole deformations is important. A general
treatment of a combined quadrupole-octupole field is proposed in
the framework of a general collective model for coupled multipole
surface modes \cite{Ro82,Ro88}.

Based on the above consideration we suggest that the most general
collective Hamiltonian of a system with octupole correlations should
contain also the standard (axial) quadrupole rotation part
\begin{equation}
\hat{H}_{rot}= A\hat{I}^{2}+A'\hat{I}_{z}^{2} \ ,
\label{Hrot}
\end{equation}
where $A$ and $A'$ are the inertial parameters.
In addition the following higher order diagonal quadrupole-octupole
interaction term (corresponding to the product $Y_{2\, 0}\cdot Y_{3\,
0}$) could be introduced:
\begin{equation}
\hat{H}_{qoc}=f_{qoc}\frac{1}{I^{2}}(15\hat{I}_{z}^{5}-
                  14\hat{I}_{z}^{3}\hat{I}^{2}+
                  3\hat{I}_{z}\hat{I}^{4}) \ .
\label{Hqoc}
\end{equation}
This operator is normalized with respect to the multiplication
factor $I^3$. (More precisely we use the product $I^3Y_{2\, 0}\cdot
Y_{3\, 0}$) so as to keep all non-quadrupole Hamiltonian terms of
the same order.)

Then the Hamiltonian of the system can be written as
\begin{equation}
\hat{H}=\hat{H}_{bh}+\hat{H}_{rot}+\hat{H}_{oct}+
\hat{H}_{qoc} \ .
\label{Hgen}
\end{equation}
Here
\begin{equation}
\hat{H}_{bh}=\hat{H}_{0}+f_{k}\hat{I}_{z} \ ,
\end{equation}
is a pure phenomenological part introduced to reproduce the bandhead
energy in the form
\begin{equation}
E_{bh}=E_{0}+f_{k}K \ ,
\end{equation}
were $E_{0}$ and $f_{k}$ are free parameters. The $K$-dependence of
$E_{bh}$, which can be reasonably referred to the intrinsic
motion, provides the correct value of the bandhead angular momentum
projection $K$ in the variation procedure described below.

We remark that the Hamiltonian (\ref{Hgen}) is not a rotational
invariant in general. It does not commute with the total angular
momentum operators and any state with given angular momentum is
energy split with respect to the quantum number $K$. Therefore, the
physical relevance of this Hamiltonian depends on the possibility to
determine in an unique way the angular momentum projection. The basic
assumption of our consideration is that K is not frozen within the
states of the collective rotational band. We suggest that for any
given angular momentum it should be determined  so as to minimize the
respective collective energy. The resulting energy spectrum
represents the yrast sequence of energy levels for our model
Hamiltonian. We remark that similar procedure is used in Refs.
\cite{HM,MQ} in reference to the $\Delta I=2$ staggering effect in
superdeformed nuclei.

As a first step in testing our Hamiltonian we consider its diagonal
part
\begin{equation}
\hat{H}^{d}=\hat{H}_{bh}+\hat{H}_{rot}+\hat{H}_{oct}^{d}+
\hat{H}_{qoc} \ .
\label{Hdiag}
\end{equation}
were the operator $\hat{H}_{oct}^{d}\equiv \hat{H}_{F_{1}(1)}$
represents the diagonal part of the pure octupole Hamiltonian
$\hat{H}_{oct}$, Eq.~(\ref{Hoctgen}).

The following diagonal matrix element is then obtained:
\begin{eqnarray}
E_{K}(I)&=&E_{0}+f_{k}K+AI(I+1)+A'K^{2}+
f_{11}\left( \frac{5}{2}K^{3}-\frac{3}{2}KI(I+1)\right)
\nonumber \\
&+&f_{qoc}\frac{1}{I^{2}}
\left( 15K^{5}-14K^{3}I(I+1)+
3KI^{2}(I+1)^{2}\right) \ .
\label{EKI}
\end{eqnarray}

Following the above assumption for the third angular momentum
projection, we determine the yrast sequence $E(I)$ after
minimizing Eq.~(\ref{EKI}) as a function of integer $K$ in the range
$-I\leq K\leq I$. The obtained energy spectrum depends on six model
parameters: $E_{0}$ essentially responsible for the bandhead energy;
$f_{k}$ which provides minimal energy for $K=K_{bh}=I_{bh}$;
$A$ and $A'$ are the quadrupole inertial parameters which
should generally correspond to the known quadrupole shapes (axes
ratios) of nuclei; $f_{11}$ and $f_{qoc}$ are the parameters of the
diagonal octupole (\ref{HF11}) and quadrupole-octupole (\ref{Hqoc})
interactions respectively. We consider the latter two parameters as
free parameters.

We applied several exemplary sets of the above parameters and
obtained the corresponding schematic energy spectra. One of them
is given in Table 1. It is seen that the ``yrast'' values of the
quantum number $K$ gradually increase with the increase of the
angular momentum $I$. We remark that they correspond to the local
minima of Eq.~(\ref{EKI}) as a function of $K$. This is
illustrated on Fig.~1. We see that these minimums are well
determined and their depth increases with the increase of the
angular momentum. Such a behavior of the spectrum corresponds to
a wobbling motion and could also be interpreted as a
multiband-crossing phenomenon. The obtained yrast sequence can be
considered as the envelope of the curves with different values of
the quantum number $K$ as it is illustrated in Fig.~2.

In addition we see that the $K$- values of the odd and the even
sequence of levels are grouped by couples which imply the presence
of odd--even staggering effect. Indeed, the presence of such an
effect is demonstrated in Fig. 3 (a)--(e), where the quantity
\begin{equation}
Stg(I)= 6\Delta E(I)-4\Delta E(I-1)-4\Delta E(I+1)+
\Delta E(I+2)+\Delta E(I-2)\ ,
\label{stag}
\end{equation}
with $\Delta E(I)=E(I+1)-E(I)$, is plotted as a function of
angular momentum $I$ for several different sets of model
parameters. (The quantity $Stg(I)$ is the discrete approximation
of the fourth derivative of the function $\Delta E(I)$, i.e. the
fifth derivative of the energy $E(I)$. Its physical relevance has
been discussed extensively in Refs \cite{DBoct00,MDRRB00}.)

Fig.~3(a) illustrates a long $\Delta I=1$ staggering pattern with
several irregularities, which looks similar to the ``beats''
observed in the octupole bands of some light actinides such as
$^{220}$Ra, $^{224}$Ra and $^{226}$Ra  \cite{DBoct00}. Also, it is
rather similar to the staggering patterns observed in rotational
spectra of diatomic molecules \cite{RDM97}. In Fig.~3(b) the
increased values of $f_{11}$ and $f_{qoc}$ provide a wide angular
momentum region (up to $I\sim 40$) with a regular staggering
pattern. The further increase of $f_{qoc}$ results in a staggering
pattern with different amplitudes, shown in Fig.~3(c). These two
figures resemble the staggering behavior of some rotational
(negative parity) bands based on octupole vibrations
\cite{NMSDunp}. The further increase of $f_{11}$ and $f_{qoc}$
leads to a staggering pattern with many ``beats'', as shown in
Fig~3(d). Notice that in Fig. 3(d) the first three ``beats'' are
completed by $I\approx 40$, while in Fig. 3(a) the first three
``beats'' are completed by $I\approx 70$. An example with almost
constant staggering amplitude is shown in Fig.~3(e). It resembles
the form of the odd--even staggering predicted in the SU(3) limit
of various algebraic models (see Ref. \cite{DBoct00} for details
and relevant references). It also resembles the odd--even
staggering seen in some octupole bands of light actinides, such
as $^{222}$Rn \cite{DBoct00}.

Now we can discuss the general Hamiltonian structure (\ref{Hgen})
including the various non-diagonal terms (\ref{HA}),
(\ref{HF12})--(\ref{HF23}). Here, the major problem is the
circumstance that $K$ is generally not a good quantum number.
However we are able to provide our analysis for small values of
the respective parameters which conserve $K$ ``asymptotically''
good. This requirement assumes a weak $K$-bandmixing interaction
which guarantees that for any explicit energy minimum appearing
in the diagonal case the corresponding perturbed Hamiltonian
eigenvalue will be uniquely determined. Thus we are able to obtain
respective $K$-mixed yrast energy sequence. Our numerical analysis
of the Hamiltonian eigenvector systems shows that the parameters
of the non-diagonal terms should be by a order smaller in value
than the parameter $f_{11}$. In addition, we established that the
following couples of non-diagonal terms give the same
contribution in the energy spectrum: $\hat{H}_{A_{2}}$ and
$\hat{H}_{F_{2}(1)}$; $\hat{H}_{F_{1}(2)}$ and
$\hat{H}_{F_{1}(3)}$; $\hat{H}_{F_{2}(2)}$ and
$\hat{H}_{F_{2}(3)}$.

In Fig.~3(f) a staggering pattern with a presence of
$K$-bandmixing is illustrated. In fact we added the following
three non-diagonal terms $\hat{H}_{F_{1}(2)}$,
$\hat{H}_{F_{2}(1)}$ and $\hat{H}_{F_{2}(2)}$ to the already
considered diagonal Hamiltonian (\ref{Hdiag}), with the
parameters of the latter being kept the same as in Fig.~3(b) (and
in Table~1). We see that the mixing leads to a decrease in the
staggering amplitude with the increase of angular momentum so
that the staggering pattern is reduced completely in the higher
spin region. This pattern  resembles the experimental situation
in $^{218}$Rn and $^{228}$Th \cite{DBoct00} (odd--even staggering
with amplitude decreasing as a function of $I$).

So, the staggering patterns illustrated so far (Fig.~3) cover
almost all known $\Delta I=1$ staggering patterns in nuclei and
molecules. The amplitudes obtained for the examined sets of
parameters vary up to $300$ keV. Some reasonable theoretical
patterns with $Stg(I)\sim 500$keV can be easily obtained. On this
basis we suppose that the model parameters can be adjusted
appropriately so as to reproduce the staggering effects in nuclear
octupole bands as well as in some rotational negative parity bands
built on octupole vibrations. Also, an application of the present
formalism to the spectra of diatomic molecules could be
reasonable.

Here the following comments on the structure of the collective
interactions used and the related symmetries would be relevant:

1) The equal contribution of the three couples of non-diagonal
terms (mentioned above) indicates that only four octupole
Hamiltonian terms are enough to determine the energy spectrum.
This result  reflects the circumstance that in the intrinsic
frame three octupole degrees of freedom, from the seven ones, are
related to the orientation angles. For example we could suggest
that the following terms [applied in Fig.~3(f)] give an
independent contribution in the total Hamiltonian:
$\hat{H}_{F_{1}(1)}$, $\hat{H}_{F_{1}(2)}$, $\hat{H}_{F_{2}(1)}$
and $\hat{H}_{F_{2}(2)}$. We remark that our analysis (related to
the collective rotations of the system) gives a natural way to
determine the four collective octupole interaction terms.

2) From symmetry point of view we remark that the diagonal term
$\hat{H}_{F_{1}(1)}$ which corresponds to $Y_{3\, 0}$ possesses an
axial symmetry while  the non-diagonal terms $\hat{H}_{F_{1}(2)}$,
$\hat{H}_{F_{2}(1)}$ and $\hat{H}_{F_{2}(2)}$ (of previous item
1)) are constructed by using the combinations $(Y_{3\, 1}-Y_{3\,
-1})$ with C$_{2v}$ symmetry, $(Y_{3\, 2}+Y_{3\, -2})$ with
T$_{d}$ symmetry and $(Y_{3\, 3}-Y_{3\, -3})$ with D$_{3h}$
symmetry. So, our analysis shows that the axial symmetric term
should play the major role in the structure of the collective
rotational Hamiltonian while the non-axial parts could be
considered as small $K$-band-mixing interactions. From
microscopic point of view, a detailed analysis of the above
spherical harmonic combinations and the respective symmetries has
been provided on the basis of the one particle spectra of the
octupole-coupled two-level model \cite{HBXZ91}.

3) The observed influence of the non-diagonal Hamiltonian terms on
the fine structure of our ``schematic'' spectra suggests an
important physical conclusion: the non-diagonal $K$- mixing
interactions suppress the staggering pattern. In such a way we
find that the axial symmetric term $\hat{H}_{F_{1}(1)}$ is the
only one pure octupole degree of freedom which provides ``beat''
staggering behavior of the quantity (\ref{stag}) (See Fig.~3(e)).
(The quadrupole--octupole term $\hat{H}_{qoc}$ gives an
additional contribution and provides wider angular momentum
regions with regular staggering.) So, our analysis suggests that
the  $\Delta I=1$ staggering effect observed in systems with
octupole deformations could be referred to the dominant role of
the axial symmetric ``pear-like'' shape.

In addition, it is important to remark that the fine (staggering)
behavior of our schematic energy spectra reflects the structure of
the interactions considered through the K- sequences generated in
the above minimization procedure. Thus our analysis suggests that
in the high angular momentum region some high K band structures
should be involved. From microscopic point of view the values
$K=0$, 1, 2, 3 have been included in calculations, showing that
in the beginning of the rare earth region the values $K=0$, 1 are
important for the lowest $3^-$ state, while in the middle of the
region the values $K=1$, 2 are important and in the far end of
the region the values $K=2$, 3 are important \cite{NV70a}. The
same authors deal with nuclei with $A\geq 222$ in Ref.
\cite{NV70a}. One of the authors of Refs. \cite{NV70a,NV70b} in
Ref. \cite{V70} finds that the restriction to $K\leq 3$ is not
justifiable for large energies. This is in agreement with our
findings of Table 1.

In conclusion, we remark that the collective interactions
considered in this work suggest the presence of various fine
rotational band structures in quantum mechanical systems with
collective octupole correlations. In particular, they provide
various forms of staggering patterns which appear as the results
of a delicate interplay between the terms of pure octupole field
and the terms of high order quadrupole--octupole interaction. The
analysis carried out outlines the dominant role of the axial
symmetric ``pear-like'' shape for the presence of a $\Delta I=1$
staggering effect. The obtained multi K- band crossing structures
could be referred to a wobbling collective motion of the system.
We propose that the interactions involved would provide a
relevant handle in the study of collective phenomena in nuclei
and other quantum mechanical systems with complex shape
correlations.

\bigskip
\noindent {\bf Acknowledgments}
\medskip

This work has been supported by the Bulgarian National Fund for
Scientific Research under contract no MU--F--02/98. We are grateful
to Prof. P. Quentin for the illuminating discussions.

\newpage

    \begin{center}
    {\bf Figure Captions}
    \end{center}
    \bigskip\bigskip

\noindent {\bf Figure 1.} The diagonal energy matrix element
$E_{K}(I)$ (in MeV), Eq.~(\protect\ref{EKI}), is plotted as a
function of $K$ for $I=1,2,...,10$, for the parameter set
$E_{0}=500$keV, $f_{k}=-7.5$keV, $A=12$keV, $A'=6.6$keV,
$f_{11}=0.56$keV, $f_{qoc}=0.085$keV.
\bigskip

\noindent {\bf Figure 2.} The diagonal energy matrix element
$E_{K}(I)$ (in MeV), Eq.~(\protect\ref{EKI}), is plotted as a
function of $I$ for $K=10,11,12,13$, for the parameter set of
Figure 1.
\bigskip

\noindent {\bf Figure 3.} $\Delta I=1$ staggering patterns
[Eq.~(\protect\ref{stag})] obtained: (a) -- (e) by the diagonal
Hamiltonian (\ref{Hdiag}) for several different sets of model
parameters; (f) by adding three non-diagonal terms
$\hat{H}_{F_{1}(2)}$ [Eq.~(\ref{HF12})], $\hat{H}_{F_{2}(1)}$
[Eq.~(\ref{HF21})] and $\hat{H}_{F_{2}(2)}$ [Eq.~(\ref{HF22})] to
the diagonal Hamiltonian (\ref{Hdiag}).
\bigskip


\begin{table}
\caption{The ``yrast'' energy levels, $E(I)$ (in KeV), and the
respective $K$- values obtained by Eq.~(\protect\ref{EKI}) for the
parameter set $E_{0}=500$keV, $f_{k}=-7.5$keV, $A=12$keV,
$A'=6.6$keV, $f_{11}=0.56$keV, $f_{qoc}=0.085$keV.}

    \bigskip
    \begin{center}
    \begin{tabular}{ccccccccc}
    \rule{0em}{2.2ex}
\\
\hline\hline
$I$&$E(I)$&$K$&$I$&$E(I)$&$K$&$I$&$E(I)$&K \\
\hline
1 & 522.772& 1&  13&  2335.81&  5 & 25&  5453.12&  11 \\
2 & 568.327& 1&  14&  2576.57&  6 & 26&  5694.49&  12 \\
3 & 637.095& 1&  15&  2827.57&  6 & 27&  5935.5 &  12 \\
4 & 728.71 & 1&  16&  3082.36&  7 & 28&  6157.5 &  13  \\
5 & 840.857& 2&  17&  3344.94&  7 & 29&  6378.29&  13  \\
6 & 971.155& 2&  18&  3608.18&  8 & 30&  6575.37&  14  \\
7 & 1123.22& 2&  19&  3877.05&  8 & 31&  6770.62&  14  \\
8 & 1288.09& 3&  20&  4143.16&  9 & 32&  6937.23&  15  \\
9 & 1472.71& 3&  21&  4413.03&  9 & 33&  7101.62&  15  \\
10& 1668.56& 4&  22&  4676.45&  10& 34&  7232.21&  16  \\
11& 1880.56& 4&  23&  4942.01&  10& 35&  7360.44&  16  \\
12& 2101.68& 5&  24&  5197.18&  11& 36&  7449.45&  17  \\
    \hline\hline
    \end{tabular}
    \end{center}
    \label{tab:spect}
    \end{table}

\ \ \ \ \ \

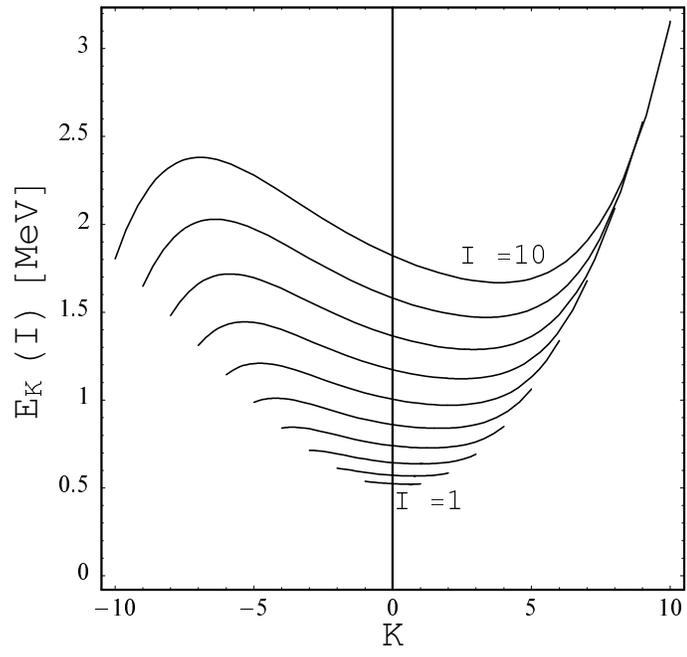
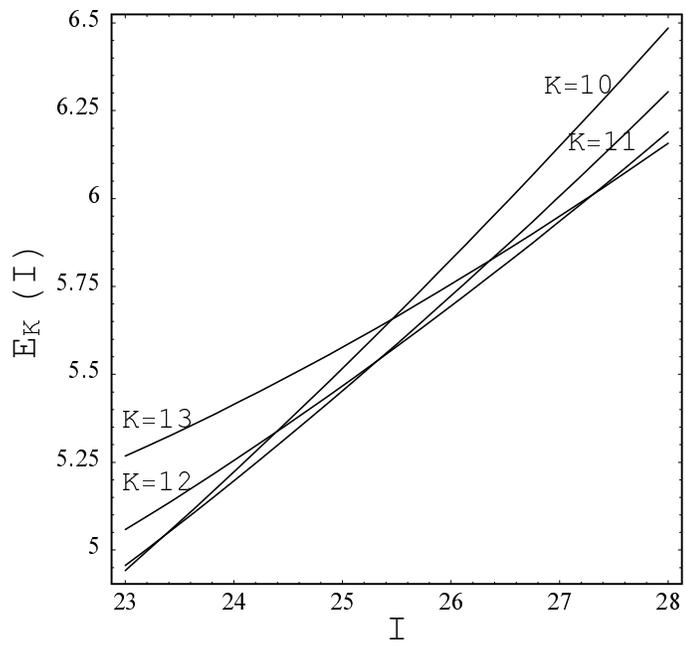
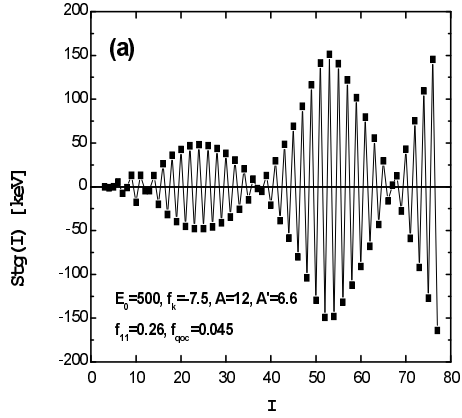
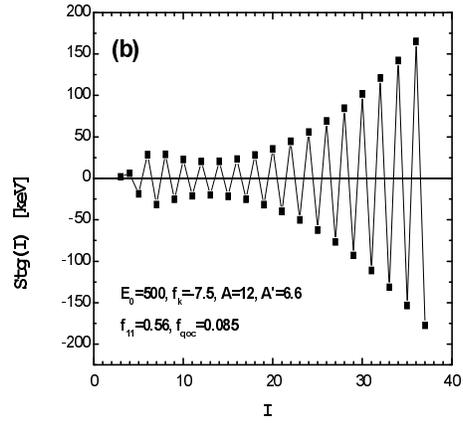
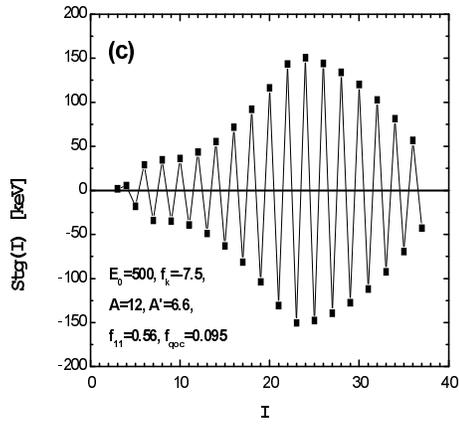
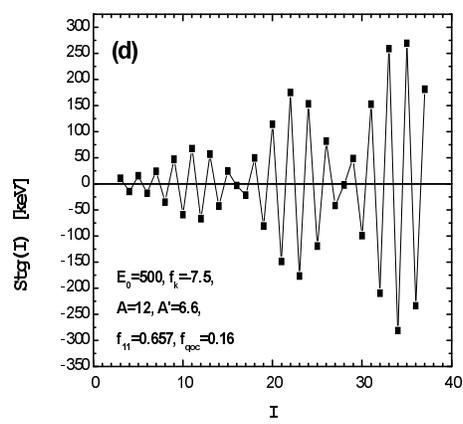
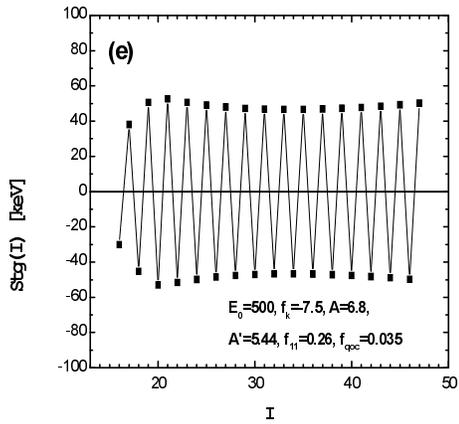
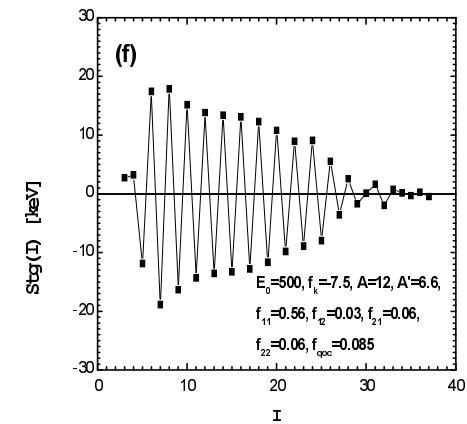
\end{document}